# Microscopic Model of Cuprate Superconductivity

Richard H. Squire $^\xi$, Norman H. March*

$^\xi$ Department of Chemistry, West Virginia University
Institute of Technology
Montgomery, WV 25303, USA

* Department of Physics, University of Antwerp (RUCA), Groenborgerlaan 171,
B-2020 Antwerp, Belgium
and
Oxford University, Oxford, England

## ABSTRACT

We present a model for cuprate superconductivity based on the identification of an experimentally detected "local superconductor" as a charge 2 fermion pairing in a circular, stationary density wave. This wave acts like a highly correlated local "boson" satisfying a modified Cooper problem with additional correlation stabilization relative to the separate right- and left-handed density waves composing it. This local "boson" could be formed in a two-bound roton-like manner; it has Fermion statistics. Delocalized superconductive pairing (superconductivity) is achieved by a Feshbach resonance of two unpaired holes (electrons) resonating with a virtual energy level of the bound pair state of the local "boson" as described by the Boson-Fermion-Gossamer (BFG) model. The spin-charge order interaction offers an explanation for the overall shape of the superconducting dome as well a microscopic basis for the cuprate $T_c\text{'}s$. An explanation of the correlation of $T_c$ with experimental inelastic neutron and electron Raman scattering is proposed, based on the energy of the virtual bound pair. These and other modifications discussed suggest a microscopic explanation for the entire cuprate superconductivity dome shape.

## 1. Introduction

Almost everyone agrees that the multitude of high $T_c$ superconducting compounds made with at least one copper-oxygen plane per unit cell can be described by a "universal" phase diagram (Fig 1) [1]. In addition general consensus seems to be achieved that these compounds are derived from a Mott insulator and might be expected to be metallic, but the intervention of strong electron-electron repulsion has restricted electron (hole) motion. Upon doping these insulators the strong antiferromagnetism is destroyed, but it appears that the strong electron-electron repulsion persists. Some of the original repulsion persists resulting in unique types of correlation. The well known stripes [2] are one form of the correlation along with electronic liquid crystals [3], and we have proposed that an additional novel type of correlation, namely a charge 2 circular density wave is also present at a certain range of doping in both electron- and hole-doped materials [4]. We expect this density wave to create a singlet state at rather high temperatures and in certain ways the formation of this entity resembles a roton. Below we develop a model containing the density wave which describes how delocalized superconducting electron pairing occurs along with a rationale for the $T_c$ variance of the families of superconductors, the variations due to interlayer substitutions and the dome-like appearance of the superconducting phase and possible quantum critical points. The foundation of this model is mostly our own work supported by others' concepts which seem to be connected in a natural manner. The final model seems to be compatible with most of the huge body of excellent experimental work.

Certainly any complete model of cuprate superconductivity will be required to explain a multitude of novel experimental results that have been accumulating since the discovery of cuprate superconductivity [5]. Instead of this daunting task, our objective in this manuscript is to focus on the model alluded to above and implications for the phase diagram of cuprate superconductivity by rationalizing what we deemed to be certain critical connections between theory and experimental results (Section 2). Section 3 contains a discussion of the possible roton resemblance, followed by an analysis of the stability of a single and paired-CDW, the latter sometimes called a "local superconductor" in the literature. We then compare and contrast a two-roton bound state with a 2-CDW bound state. Section 4 discusses the cuprate phase diagram beginning with the underdoped pairing, followed by the essential components of a Feshbach resonance and finishing with competing orders in the overdoped portion to finish the phase diagram. Section 5 contains a summary, conclusions and future work and the Appendix discusses phase stiffness and ODLRO.

## 2. Initial Development of the Model Together with Some Key Experiments.

### A. Previous Theoretical and Experimental Work.

The doping of two electrons on a fulleride model should according to MO/band theory produce a conductor; instead an insulator is formed. Building on previous work [6], it seemed that a Jahn-Teller distortion could generate a harmonic well around the equator of a fulleride molecule, a "trap" if you will similar in spirit to traps generated in "cold atom" experiments. There, certain alkali molecules have been shown to undergo condensation to form bosons. Of course there are many differences between these experiments and fulleride superconductivity, but the notion of a harmonic trap lead us to consider that perhaps the two doped electrons had become much more highly correlated and more tightly bound than a typical Jahn-Teller stabilization. We proposed the enhanced stability might be accomplished by means of a circular Peierls distortion containing two electrons moving in opposite direction around a fulleride molecule as if it were a "localized" Cooper pair (or "localized superconductor").

In "pre-BCS" times Peierls proposed a theorem [7] which states that a 1D equally spaced chain with one electron per ion is unstable. In one dimension, if each ion contributes one electron, the energy band will be half filled to values of $k = \pm \pi/2a$, where a is the lattice constant. Peierls also recognized that the lattice is unstable to distortion as every other ion moves closer to one neighbor and away from the other with the result that the period has doubled to 2a and speculated that this distortion causes a small energy savings which should be noticeable at lower temperatures. This distortion has been thought early on to be the force responsible for bond alteration in certain chemical compounds [8]. Peierls recognized the similarities of this phenomenon to superconductivity, i.e. an energy gap opening at the Fermi level, but much later, when the organic materials TTF-TCNQ were examined in the 1970's, this effect instead produced an insulating transition, as in the fullerides. Indeed, Frohlich's adaptation [9] of Peierls model has been used to propose a different version of superconductivity in organic superconductors that arises from coupling of the electrons with a moving charge density lattice wave instead of a BCS-type pairing [10].

Later, Su-Schrieffer-Heeger (SSH) [11] used the Peierls' distortion as the underlying basis to describe superconductivity in polyacetylene (PA). They recognized that the reorganization energy described above would produce a double well minimum from bond alteration which is associated with spontaneous symmetry breaking resulting in a two fold degenerate state. On doping, this system permits nonlinear excitations which are equivalent to moving domain walls which act as topological solitons and give rise to unusual charge and spin states.

We viewed our model as an extension of SSH by adding a second fermion which would result in a charge 2e density wave [12]. However, we agree with others [13] that instead of the copper oxide planar system maintaining a linear density wave structure, the system has enhanced fluctuations (which some have called "magnetic roton" excited states) because of the proximity to a phase with a charge order and a propensity to form a roton as in helium and certain atomic BEC systems. The result is an order comparable in energy and a Bragg peak will begin forming and produce a spike in the structure factor, as may have been observed [13]. We have discussed another such possible charge order, a Wigner crystal, in the fullerides [14]. However, in the fullerides since two electrons are known to be involved, we have considered the correlated structure to be the "localized" superconductor discussed above with most features exactly the same as a superconductor, gap, etc. However, there is one major difference, namely, a **local** off-diagonal order (called by some Off-Diagonal Short Range Order –ODSRO) in the density matrix instead of long range order. The result is a "Cooper pair" insulator, (see [15]).

We now present two pieces of experimental evidence that could substantiate our proposed model: The first piece of experimental evidence that we want to use is the particularly informative STM data of Kohsaka et al [16], who have provided tantalizing glimpses of cuprate superconductivity in certain scanning tunneling microscope (STM) experiments. Specifically, in $Ba_2Sr_2CaCu_2O_{8+\delta}$, they use quasi-particle interference (QPI) imaging that allows them to determining the material's electronic structure in k- and r-space simultaneously. These experiments show that the Bogoliubov quasi-particle excitations are confined to a specific part of k-space known as the "Bogoliubov arc" that represents regions of k-space containing delocalized Cooper pairs as doping decreases. Adding a few holes to the antiferromagnetic ground state of a cuprate Mott insulator transforms the system into a superconductor via delocalized Cooper pairs in k-space. When this transformation happens, two types of electronic excitations appear; one, a pseudogap at higher energy, and lower energy Bogoliubov quasiparticles from the breakup of Cooper pairs. Through the simultaneous imaging of both r-space and k-space, the Bogoliubov quasiparticle region shrinks rapidly while transferring spectral weight to the higher energy r-space states which are localized Cooper pairs, breaking translational and rotational symmetries at the atomic scale; these are the pseudogap states. Delocalized Cooper pairs vanish as the Mott insulator is approached, leaving "localized" Cooper pairs. Our model can describe this.

The second piece of evidence is that a) fulleride superconductivity is similar in some respects to the cuprates based on the Uemura muon spin relaxation work plot of $T_c$ vs. the effective Fermi temperature, derived from the response of various superconducting systems superfluid response [17]. b) We had proposed a charge 2 circular density wave in the fullerides, and the possibility that a roton-like entity could be generated by a transverse interaction of two charge density waves (or vibrations/phonons in a molecular case), especially if the system were in proximity to some type of charge ordering, seemed to support our contention that the fullerides and cuprates might have a similar superconducting mechanism. Others [13, 18, 19, 20, 21] have also suggested that a "magnetic roton" formation, similar to helium, the fractional Quantum Hall Effect, and other systems [22], might explain the correlation of the neutron and Raman scattering since the former detects a triplet in the spin resonance while the latter operates in the charge channel (Fig 2).

How exactly this information relates to a superconducting pairing mechanism or any additional mechanistic features of HTSC was not initial apparent to us. We recently proposed a charge-spin complex [4c] that is expected to be different than a Landau-Feynman roton because superconductors carry charge (our model a charge 2 entity), have spin and charge interactions, are paired Fermions, have different potential energy surfaces, and we believe contain a Feshbach resonance pairing, etc. However, we recognize that the neutron and Raman scattering mentioned could relate to the same type of charge-spin complex. Is the formation due to a Jahn-Teller interaction or is it possible that each of the two interacting circular charge density waves might be formed in a manner analogous to a roton? Both can provide spontaneous symmetry breaking, so certainly further investigation was warranted to uncover details of what we propose is the low energy excitations of the system. The purpose of this manuscript is to investigate the various possibilities.

Lastly, the seminal work of Nozieres and Schmidt-Rink's (NSR) examination of the crossover from BCS to BEC was the first mention of what has come to be known as a preformed pair, a boson-like object formed by two fermions [23]. NSR recognized that this preformed pair would be considerably smaller than a BCS Cooper pair, and more tightly bound (Fig 3). The exact origin of the pair has been a mystery with several proposed candidates ranging from "electron lone pairs" in chemical parlance at one extreme to various assumptions to be justified by the results obtained at the other extreme. The importance of the "local superconducting pair" is that it can be identified as the "boson" in the fermion-boson resonance model [24] which we have discussed previously [4]. Others have described how a linear CDW competes with BCS superconductivity, but in the high temperature SC cuprates, the localized Cooper pairs (or preformed pairs) we have described are essential as they provide the crucial component which sets the energy scale in the formation of delocalized Cooper pairs. This determines $T_c$ for the various cuprate materials as well as provides the leading competing order in the cuprates (and possibly other materials') superconductivity.

B. **Rotons.**

The LSC's we are proposing seem to have more features than a roton, but seemingly similar origins, namely a "softening" of a linear phonon mode (Fig 4). Landau proposed the existence of a roton a number of years ago as the most important elementary excitation in the helium phonon spectrum [25]. It is a minimum in the otherwise linear phonon dispersion mode as observed by neutron scattering which is caused by a softening of the phonon mode as solidification of $^4$He approaches but has not yet been achieved. The roton is accompanied by a peak in the structure factor density matrix at the same wave vector and the two are related by a Fourier transform. Alternatively, constructed low temperature atomic BEC systems that are not close to, say, a Mott transition or other type of charge ordering, will probably not exhibit these features. The Wigner crystal, proposed in the fullerides [14] and suggested in the hole doped Cuprates, specifically the 4 x 4 unit cells seen in the STM spectra [13], are a type of ordering where rotons created by density (wave) fluctuations might exist.

Landau's form of the Hamiltonian describing many particle systems which could be a quantum liquid or solid depending on the interaction is [26]

$$H = \int \left( m \frac{\vec{v} \bullet \rho \vec{v}}{2} \right) d^3r + U(\rho) \qquad (1)$$

where $\frac{\vec{v} \cdot \rho \vec{v}}{2}$ is the kinetic energy operator and the potential energy operator $U(\rho)$ is a functional of the density operator, $\rho(\vec{r}) = \sum_i \delta(\vec{r} - \vec{r}_i)$, $\vec{r}_i$, the position of the ith particle. Also in first quantized notation, the velocity operator can be expressed as

$$\vec{v}(\vec{r}) = \sum_i \left[ \frac{\vec{p}}{2m} \delta(\vec{r} - \vec{r}_i) + \delta(\vec{r} - \vec{r}_i) \frac{\vec{p}_i}{2m} \right]$$

The commutation relation between the two is

$$[\rho(\vec{r}'), \vec{v}(\vec{r})] = i \frac{\hbar}{m} \nabla \delta(\vec{r}' - \vec{r}) \qquad (2)$$

We examine a liquid material with uniform density, but modulated by a lattice, to understand the density excitation spectrum where the average velocity is zero, following others [18, 19, 25, 26]. The Fourier expansions of the density and velocity operators are

$$\rho(\vec{r}) = \rho_0 + \frac{1}{N} \sum_{\vec{k}} \vec{\rho}_{\vec{k}} e^{i\vec{k} \cdot \vec{r}}$$

$$\vec{v}(\vec{r}) = \frac{1}{N} \sum_{\vec{k}} \vec{v}_{\vec{k}} e^{i\vec{k} \cdot \vec{r}}$$

with lattice normalization as the number of sites, N, instead of the volume, V. The velocity in a superfluid is proportional to the gradient of the local phase of the particle, so in Fourier components this is

$$\vec{v}_k = \frac{\hbar}{m} \nabla \theta \Rightarrow \vec{v}_{\vec{k}} = \frac{i\hbar \vec{k}}{m} \theta_{\vec{k}}$$

and from (2), the density phase commutation relationship (or number phase commutation, see Appendix), i.e. conjugate variables,

$$[\rho_{\vec{k}}, \theta_{i,-\vec{k}'}] = i \delta_{\vec{k}\vec{k}'}$$

Substituting this into (1),

$$H = U(\rho_0) + \frac{1}{N} \sum_{\vec{k}} \left[ \frac{\rho_0 \hbar^2 k^2}{2m} |\theta_{-\vec{k}}|^2 + \frac{1}{2} \left( U_{\vec{k}} + \frac{\hbar^2 k}{4m\rho_0} \right) |\rho_{\vec{k}}|^2 \right]$$

with $U_{\vec{k}}$, the Fourier transform of the interaction. Here, the quantum liquid is a collection in momentum space of harmonic oscillators for the density fluctuations, and the last term keeps track of the phase and density degrees of freedom. Accordingly, the spring constant and mass of the oscillators is, respectively,

$$M_{\vec{k}} = \frac{m}{\rho_0 k^2} \qquad\qquad K_{\vec{k}} = U_{\vec{k}} + \frac{\hbar^2 k^2}{4m\rho_0}$$

The ground state of the quantum liquid is $U(\rho_0) + \sum_{\vec{k}} \hbar \omega_{\vec{k}} / 2$. The quantum liquid excitation spectrum is

$$E_{\vec{k}} = \hbar \omega_{\vec{k}} \left( n + \frac{1}{2} \right) \text{ where } \omega_{\vec{k}}^2 = \frac{K_{\vec{k}}}{M_{\vec{k}}} = \frac{\rho_0 k^2}{m} \left( U_{\vec{k}} + \frac{\hbar^2 k^2}{4m\rho_0} \right)$$

The Fourier transform of the density-density correlation function is the structure factor, $S(\vec{k})$ which we define as,

$$S(\vec{k}) = \frac{\langle |\rho_{\vec{k}}|^2 \rangle}{N\rho_0} \qquad (3)$$

Using the Virial theorem,

$$\frac{1}{2}\hbar\omega_{\vec{k}} = \left(U_{\vec{k}} + \frac{\hbar^2 k^2}{4m\rho_0}\right)\langle |\rho_{\vec{k}}|^2 \rangle$$

Then,

$$\hbar\omega_{\vec{k}} = \frac{\hbar^2 k^2}{2mS(\vec{k})}$$

which is the same result as Feynman obtained for helium except we have an expression for the contributions in terms of harmonic oscillators which can be related to our density wave expressions.

The structure factor for the quantum liquid we are discussing is then

$$S(\vec{k}) = \frac{\hbar k}{2m} \frac{1}{\sqrt{\frac{\rho_0}{m}\left(U_{\vec{k}} + \frac{\hbar^2 k^2}{4m\rho_0}\right)}}$$

The ground state, $\psi_{gs}$, will only contain the zero point energy of the harmonic oscillators. Following others [18, 19, 25, 26] and using the RPA [27], we can write a ground state wave function with density correlations as

$$|\Psi\rangle = \exp\left[-\frac{1}{2}\sum_{\vec{k}} \sqrt{\frac{m}{\rho_0 \hbar^2 k^2}\left(U_{\vec{k}} + \frac{\hbar^2 k}{4m\rho_0}\right)}\right]|\psi_{gs}\rangle$$

which is correct to quadratic order. It is noteworthy that the exponential factor suppresses density fluctuations. But as spontaneous crystallization approaches, there will be enhanced density fluctuations in the electron fluid. The roton appears as a damped pole in the density-density correlation function, and since the structure factor is the Fourier transform, the roton will appear as a peak of the structure factor vs. the momentum in three dimensions.

**C. Charge Density Waves [28].**

In the most general case a charge density wave can be examined in the adiabatic limit assume the electrons follow the ions instantaneously by using the electron-phonon (Fröhlich) Hamiltonian,

$$H = \sum_k \varepsilon_k a_k^\dagger a_k + \sum_q \hbar\omega b_q^\dagger b_q + \sum_{k,q} g_q a_{k+q}^\dagger a_k \left(b_{-q}^\dagger + b_q\right)$$

where a and b are electron (or hole) and phonon operators, respectively, with momentum k and q, $g_q$ is the coupling constant and $\varepsilon$ and $\omega$ are the energies. The effect on the normal vibration coordinates Q for small amplitude displacement is expressed as

$$\hbar^2 \ddot{Q}_q = -\left[\left[Q_q, H\right], H\right]$$

Since the commutator $\left[Q_q, P_{q'}\right] = i\hbar \delta_{q,q'}$, we have

$$\ddot{Q}_q = -\omega_q^2 Q_q - g\left(\frac{2\omega_q}{M\hbar}\right)^{1/2} \rho_q$$

with $\rho_q = \sum_k a_{k+q}^\dagger a_k$ being the q$^{th}$ component of the electron density (g is assumed independent of q). The second term on the RHS is an effective force constant due to combined electron-vibration interaction and the ionic potential $g\left(\frac{2M\omega_q}{\hbar}\right)^{1/2} Q_q$ results in a density fluctuation

$$\rho_q = \chi(q,T) g\left(\frac{2M\omega_q}{\hbar}\right)^{1/2} Q_q$$

Using linear response theory leads to the following equation of motion

$$\ddot{Q}_q = -\left[\omega_q^2 + \frac{2g^2\omega_q}{M\hbar}\chi(q,T)\right]Q_q$$

which produces a renormalized vibration frequency

$$\omega_{ren,q}^2 = \omega_q^2 + \frac{2g^2\omega_g}{M\hbar}$$

In a 1D model $\chi(q,T)$ has its maximum value at $q = 2k_F$, the so-called Kohn anomaly (Fig 4) where the reduction (softening) of the vibrational frequency is most significant. Here

$$\omega_{ren,2k_F}^2 = \omega_{2k_F}^2 - \frac{2g^2 n(\varepsilon_F)\omega_{2k_F}}{\hbar}\ln\left(\frac{1.14\varepsilon_0}{k_B T}\right)$$

As the temperature is reduced, the renormalized vibration frequency goes to zero which defines a transition temperature where a frozen-in distortion occurs. From the equation above a mean field transition temperature for a charge density wave, $T_{CDW}^{MF}$, can be calculated

$$k_B T_{CDW}^{MF} = 1.14\varepsilon_0 e^{-1/\lambda} \qquad (4)$$

with $\lambda$, the electron-vibration coupling constant (dimensionless) defined as

$$\lambda = \frac{g^2 n(\varepsilon_F)}{\hbar \omega_{2kF}}$$

This resulting gap equation is **exactly identical to the BCS equation**, but there is no off-diagonal long range order (ODLRO).

## 3. Comparing and Contrasting "Localized Superconductivity" (LSC) with Rotons.
### A. "Localized" Superconductivity.

We have suggested that circular density waves are generated on a fulleride molecule by doping the proper level of electrons. If only one electron on average is doped, a single charge density wave can have certain solitons-like characteristics. However, two doped, interacting electrons on a fulleride molecule can interact with molecular vibrations and we propose that the resulting interaction can result in two charge density waves moving in a highly correlated manner in opposite directions around the molecule. The result is a large energy stabilization such that "pairing" between the electron charge density waves takes place (essentially a standing wave). This reasoning provides an explanation as to why a singlet insulator is experimentally observed instead of a metal. We suggest that this phenomenon has been experimentally observed elsewhere and described as "localized" superconductivity (LSC).

Below the phase transition we are proposing that the collective renormalized vibration frequency is zero since the lattice distortion is "frozen" as a molecular vibration mode with expectation values $\langle b_{2k_F} \rangle = \langle b^{\dagger}_{-2k_F} \rangle \neq 0$ containing two highly correlated electrons (or holes). Defining an order parameter for this charge 2 entity,

$$|\Delta|e^{i\phi} = g\left(\langle b_{2k_F} \rangle + \langle b^{\dagger}_{-2k_F} \rangle\right) \tag{5a}$$

The lattice displacement can now be shown to be

$$\langle u(x) \rangle = \left(\frac{\hbar}{2NM\omega_{2k_F}}\right)^{1/2} \left\{ i\left(\langle b_{2k_F}\rangle + \langle b^{\dagger}_{-2k_F}\rangle\right)e^{i2k_F} + cc \right\} \tag{5b}$$

$$= \left(\frac{\hbar}{2NM\omega_{2k_F}}\right)^{1/2} \frac{2\Delta}{g}\cos(2k_F x + \phi) \tag{5c}$$

$$= \Delta u \cos(2k_F x + \phi) \tag{5d}$$

along with

$$\Delta u = \left(\frac{2\hbar}{NM\omega_{2k_F}}\right)^{1/2} \frac{|\Delta|}{g}$$

The Fröhlich Hamiltonian gets modified to

$$H = \sum_k \varepsilon_k a_k^{\dagger} a_k + \sum_q \hbar\omega b_q^{\dagger} b_q + \sum_{k,q} g_q a_{k+q}^{\dagger} a_k \langle b^{\dagger}_{-q} + b_q \rangle \tag{6}$$

and since $q = \pm 2k_F$ and $\langle b_{2k_F} \rangle = \langle b^{\dagger}_{-2k_F} \rangle$

$$H = \sum_k \varepsilon_k a_k^{\dagger} a_k + 2g\sum_k \left[ a_{k+2k_F}^{\dagger} a_k \langle b^{\dagger}_{-2k_F}\rangle + a_{k-2k_F}^{\dagger} a_k \langle b_{-2k_F}\rangle \right] + 2\hbar\omega_{2k_F}\langle b_{2k_F}\rangle^2 \tag{7}$$

The order parameter defined earlier (eq 5a) is now

$$H_{el} = \sum_k \left[ \varepsilon_k a_k^\dagger a_k + |\Delta| e^{i\phi} a_{k+2k_F}^\dagger a_k + |\Delta| e^{-i\phi} a_{k-2k_F}^\dagger a_k \right]$$

Skipping a number of steps similar to discussion above, it can be shown that there are two terms leading to this lowering of energy: 1) an electronic term

$$E_{el} = n(\varepsilon_F) \left[ -\frac{\Delta^2}{2} - \Delta^2 \log\left(\frac{2\varepsilon_F}{\Delta}\right) \right] + ...$$

and 2) a lattice term

$$E_{latt} = \frac{N}{2} M \omega_{2k_F}^2 \langle u(x) \rangle^2 = \frac{\hbar \omega_{2k_F} \Delta^2}{2g^2} = \frac{\Delta^2 n(\varepsilon_F)}{\lambda}$$

where $\lambda$ was defined earlier. The total energy change then is

$$E = E_{el} + E_{latt} = n(\varepsilon_F) \left[ -\frac{\Delta^2}{2} - \Delta^2 \log\left(\frac{2\varepsilon_F}{\Delta}\right) + \frac{\Delta^2}{2\lambda} \right]$$

For $\lambda \ll 1$ and minimizing the total energy gives

$$\Delta = 2\varepsilon_F e^{-1/\lambda} \qquad (8)$$

and a condensation (correlation) energy of

$$E_{cond} = E_{norm} - E_{CDW} = \frac{n(\varepsilon_F)}{2} \Delta^2 \qquad (9)$$

There is more correlation energy than one might suspect from two circular charge 2 density wave interactions, further lowering the energy for the two bound particles. This makes the preformed pair binding energy larger than two separate density waves and the stabilization energy considerably larger than a "traditional" Cooper pair, while the size is much smaller (Fig 3)

**B. Comparison of a LSC with a Two-Roton Bound State.**

Our initial comparison of a LSC with a "Landau roton" was awkward until we recognized that rotons can bind together in a two-roton bound state [29]. Just as the charge density waves in Section 3A, the experimentally measured two roton bound state energy is more stable than twice the binding energy of a single roton at the same temperature indicating additional correlation stabilization energy. Now, the analogy between the circular charge density waves comprising the LSC and two mass density rotons in a bound state is more understandable; this also supports our notion that both the LSC and two-roton state are a highly correlated many-body states. Are they essentially the same as regards a two-bound state? A comparison of the potentials, both essentially due to Landau, is interesting (Fig 5). As regards "paired" electrons, the tunneling properties might be different suggestion that these are two different entities. However, in terms of a Feshbach resonance, both seem suitable.

Fu et al have discussed two examples of STM studies that support the notion that the approach to a solid or other Coulomb stabilizing insulating phase with charge order is needed to lead to a "magnetic" roton minimum in the excitation spectrum [13]. In contrasting Bi2212 with NaCCOC, the STM tunneling conductance exhibits a checkerboard spatial dependence with Fourier peaks at $\vec{Q} = \left(\pm 2\pi/\lambda, 0\right), \left(0, \pm 2\pi/\lambda\right)$ with $\lambda \approx 4.5a$ for the former and $\lambda \approx 4a$ for the latter. It is suggested that NaCCOC has developed true charge order (speculated to be a soliton consisting of a charged void and a spin) while Bi2212 has incipient charge order because the charged order has not "condensed". As mentioned, the roton appears as a damped pole in the density-density correlation function, and since the structure factor is the Fourier transform, the roton will appear as a peak of the structure factor vs. the momentum in three dimensions. In one or two dimensions the situation is less clear. The Mermin-Wagner theorem states that long range order is impossible in these reduce dimensions; but a roton (or two-roton state or LSC) does not have long range order. In addition their low concentration may make detection difficult. Others have detected peaks which might correspond to LSC [13]; also the STM image of Kohsaka et al [16] seems to be a direct observation. It is interesting that earlier papers focusing on a variation of the negative-U model for fullerides found that the off-diagonal terms enhance pair binding, though not enough to offset the suppression due to the density-density term [30], which generates the roton.

Overall the LSC concept could emerge from five separate theories: 1) the SSH theory of polyacetylene where addition of a second doping entity might permit interaction between the PA chains such that a circular highly correlated LSC occur; 2) the two-roton bound state which also suggests 3) that stripes could be involved in forming a circular LSC [2]; early work on CDW by Solyom and Horowitz on charge density waves [31]; From Cooper [32] and Peierls' [7] original work; or lastly, consider the work of Fisher et al [33]. Our speculation is that this type of structure could be quite ubiquitous, just difficult to detect.

### 4. Overall Shape of the SC Dome
### A. Localized Pairing.

A singlet pairing operator has been described for an extended one dimensional (1D) electron gas as [34]
$$P^\dagger = \psi^\dagger_{1\uparrow}\psi^\dagger_{2\downarrow} - \psi^\dagger_{1\downarrow}\psi^\dagger_{2\uparrow}$$
Where $\psi^\dagger_{i\sigma}$ creates a right-going $(i=1)$ or left-going $(i=2)$ fermion with spin $\sigma$. In one dimension these operators may be expressed in terms of Bose fields and their conjugate momentum corresponding to charge and spin, respectively. Expressing this operator in the language of Bose fields, specifically the charge and spin, the pairing operator becomes
$$P^\dagger \sim e^{i\sqrt{2\pi}\Theta_{c2}} \cos\left(\sqrt{2\pi}\varphi_s\right)$$

Typically in 1D, the phase (superconducting) of the pairing operator depends on the charge fields while the amplitude is a property of the spin degree of freedom (DOF). In the original Emery/Kivelson model the system acquires a spin gap through a finite value of the cosine amplitude term, and superconductivity then appears when the charge DOF becomes phase coherent [2].

Our four hole (or electron) model is based on two charge and two spin density waves

$$P^\dagger P^\dagger \sim (e^{i\sqrt{2\pi}\Theta_{c1}}\cos\varphi_{s1})(e^{i\sqrt{2\pi}\Theta_{c2}}\cos\varphi_{s2}) = (e^{i\sqrt{2\pi}(\Theta_{c1}+\Theta_{c2})})\{cos\sqrt{2\pi}(\varphi_{s1}+\varphi_{s2})\} \quad (10)$$

We propose that the two CDW's can couple in a highly correlated manner resulting in a "localized superconductor" (LSC). This is opposite to the order of occurrence in stripes where a spin gap is first opened. Replacing the charge term from eq (10) with the expression for a LSC then reduces the superconducting transition to the resonance coupling of two spin waves (see Fig 6). It seems that this type of entity has been known for some time, although we believe our microscopic proposal is new.

A key to understanding of the cuprates superconductivity is to identify the elementary excitation participating in the superconducting transition. Certainly at very low doping in the superconducting dome, it seems that there is general agreement that the only important elementary excitations are the nodal quasiparticle. When the charge order is commensurate with the underlying lattice, the antinodal quasiparticles are strongly scattered by the two-roton $\vec{Q}_{roton}$, and since $\vec{Q}_{roton}$ does not connect the nodes, the nodal quasiparticles remain gapless assuming the charge does not get excessively strong.

To describe the nodal quasiparticles, a measurement of the London penetration depth as a function of temperature $\lambda(T)$ can be converted into the superfluid density as a function of temperature $n_s(T)$ using [17]

$$\lambda(T) = \left(\frac{m_e}{\mu_0 n_s(T) e^2}\right)^{1/2}$$

We can then measure the superfluid density as we vary temperature by

$$n_s(T) = n_s(0) - n_n(T) \quad (11)$$

The BCS theory has a finite gap function $2|\Delta|$ where the probability of exciting quasiparticles is quite low if $k_B T \ll 2|\Delta|$, since $n_n(T) \sim e^{-2\Delta/k_B T}$. But in a d-wave cuprate superconductor $\Delta \to 0$ at nodal points on the Fermi surface, so it is always possible to excite quasiparticles with the result that $n_n(T) \sim T$ as experimentally verified [35]. Thus, the nodal quasiparticles shown to lead to a linear relationship of the superfluid stiffness with the transition on the underdoped side of the dome,

$$n_s(T) = n_s(0) - aT$$

The superfluid density is small in the cuprates, leading to the suggestion that the phase fluctuation of the superconducting order parameter is the controlling factor in the underdoped portion of the transition. The phase stiffness can be represented by the Hamiltonian (see Appendix),

$$H = \frac{1}{2} n_s \int d^2\vec{r} \left(\vec{\nabla}\varphi + q\vec{A}\right)^2 \quad (12)$$

**B. Feshbach Resonance [4] is the strongest order and sets the Transition temperature.**

We can describe the delocalized SC pairing using the following truncated Hamiltonian, and terminology familiar in cold atom studies,

$$H = \sum_{\bar{p},\sigma} \varepsilon_{\bar{p}} c^{\dagger}_{\bar{p}\sigma} c_{\bar{p}\sigma} - U \sum_{\bar{p},\bar{p}'} c^{\dagger}_{\bar{p}+\bar{q}/2,\uparrow} c^{\dagger}_{-\bar{p}+\bar{q}/2,\downarrow} c_{-\bar{p}'+\bar{q}/2,\downarrow} c_{\bar{p}'+\bar{q}/2,\uparrow} \quad (13)$$

$$+ \sum_{q} \left( E^{0}_{\bar{q}} + 2\nu \right) b^{\dagger}_{\bar{q}} b_{\bar{q}} + g_r \sum_{\bar{p},\bar{q}} [b^{\dagger}_{\bar{q}} c_{-\bar{p}+\bar{q}/2,\downarrow} c_{\bar{p}+\bar{q}/2,\uparrow} + h.c.]$$

Here $c_{\bar{p}\sigma}$ and $b_{\bar{q}}$ represent the annihilation operators of a fermion with kinetic energy $\varepsilon_{\bar{p}} = p^2/2m$ and a quasi-boson (charge 2e circular density wave) with the energy spectrum $E^{0}_{\bar{q}} + 2\nu = q^2/2M + 2\nu$, respectively. In the second term $-U < 0$ is the standard BCS theory attractive interaction from non-resonant processes (we are assuming the charge 2 density wave has been formed). The threshold energy of a general composite Bose particle energy band is denoted by $2\nu$ in the third term with $2\nu = \varepsilon_{res}$, the bound state energy of the charge 2 circular density wave which sets the largest energy scale of the binding of the delocalized (superconducting) Cooper pair and thus, $T_c$. The last term is the Feshbach resonance (coupling constant $g_r$) that describes how a Cooper pair (CP) can dissociate into two fermions, or how two fermions can bind into a CP.

When a superfluid phase transition occur, the following modified "gap" equation now applies for $T_c$,

$$1 = \left( U + g_r^2 \frac{1}{2\nu - 2\mu} \right) \sum_{\bar{p}} \frac{\tanh\left( \varepsilon_{\bar{p}} - \mu \right)/2T_c}{2\varepsilon_{\bar{p}} - 2\mu} \quad (14)$$

In eq (15) $g_r^2/(2\nu - 2\mu)$ is the dominant pairing interaction mediated by a boson which becomes very large when $2\mu \to 2\nu$.

There is experimental support for the Feshbach resonance utilizing the resonance complex proposed earlier [4]. A correlation of inelastic neutron scattering (INS) in the spin channel and electronic Raman scattering (ERS) of the $A_{1g}$ mode in the charge channel can be made (Fig 2). Now the fact that a quantitative scaling correlation

$$k_B T_c = \frac{E_g}{\kappa} \text{ where } \kappa \sim 5-6 \quad (15)$$

can be made is remarkable [20]. One interpretation of this result is that the excitations participating in the superconducting transition are two degenerate modes, one in the singlet charge channel (ERS) and the other in the triplet (INS) channel, two attributes of the same two-roton charge-spin structure described previously. The charge portion is a local "boson" with binding energy $\varepsilon_{res}$ comprising the main component of the energy experimentally measured by the ERS/INS experiments (Fig 2) which determines $T_c$. This is the fundamental spin-charge interaction (Feshbach resonance) in the superconducting pairing mechanism described in the next section which we described earlier in an intuitive manner [4b and Fig 6].

It seems that the second strongest energy factor in setting $T_c$ in a particular class of cuprate superconductors is the ability of the negative-U centers in charge reservoir layers to increase the depth of the binding of the principle circular charge 2 density waves/rotons and thereby strengthen the binding of the superconducting Cooper pairs and increase $T_c$. The variation in $T_c$ from this effect in conjunction with the effect of some additional layers is well-documented [36], both the enhancements and reductions to $T_c$. In addition, there also may be a component from Josephson coupling [37], though smaller, so the overall pairing energy appears as,

$$\left[\frac{g^2}{2(\mu-\nu)}\right] + J\cos(\varphi_i - \varphi_j) \text{ with } g = \varepsilon_{FR} + \varepsilon_{negU} + ... \qquad (16)$$

Presumably the resonance creates a delocalized Cooper pair which is transported an average of $\xi$, the coherence length, then decays (in the BEC limit) with amplitude for the specific transition, $b \rightarrow 2e$ [24b, 38, 39]

$$f_{if} = \frac{\Gamma}{\mu - \nu + i\Gamma} \qquad (17)$$

and a resonance width in the BEC limit of

$$\Gamma = \left(\frac{g^2}{\pi}\right) m^{3/2} \sqrt{\nu/2} |\mu_k|^2$$

at $k = (2m\nu)^{1/2}$. It does seem possible that new Cooper pairs could be formed from the components of previous pairs (if dissipative processes are suppressed), thereby creating a mechanism for a "resonating valence bond" scenario.

The cuprate materials have been described a "negative U centers" and considerable effort has been made to develop a theory of superconductivity based on this concept alone [30, 36, 40]. The negative U concept seems to support the model presented here.

**C. Overdoped Cuprate Superconductors.**

The class of cuprate superconductors show two common features: a bell-shaped curve of the superconducting transition temperature, $T_c$, as a function of doping and a dependence of $T_c$ on the number of copper oxide layers n within a homologous series. There are a number of explanations as to why this is so: competing orders, especially a d-density wave, the pseudogap, interlayer tunneling, etc. [41], the so-called Gossamer model. Certain of these have been questioned by experiment, but even in their apparent rejection, they have stimulated the notion that certainly unusual, possibly novel approaches are required. We propose a general mechanism by which both feature arise in a universal manner; this mechanism may be augmented by other proposal but not in a necessarily uniform manner.

Certainly one possibility for an effect of layers was the original proposal of Anderson [37] that tunneling between layers might be the sole mechanism for setting $T_c$. Experiment showed that this mechanism could only explain the condensation energy in a few compounds; this is not the only one, and may even only be an enhancement. We view our local boson (pseudogap) as an energetically favorable state for low energy holes (electrons) in the normal state and suggest these holes do not tunnel in a coherent manner along the c axis (perpendicular to the copper oxide planes). This reduced kinetic energy is recovered in the superconducting state, resulting in enhanced pairing. Single-particle tunneling is irrelevant in a renormalization group argument at energies below the gap and should not enter any macroscopic arguments. Another feature in this recovery is the charge reservoir pair selection. Following Gaballe et al [36] mercury also allows enhanced pairing.

Chakravarty et al [42] have studied coupling between the layers using a free energy functional. They find that the dome shape is a result of two factors: 1) **interlayer coupling** captures the dependence of the transition temperature which rises up to an optimum of n = 3, then a large doping imbalance as n>3 between the layers combined with 2) **completing order** reducing $T_c$. When n = 1 the free energy model can be solved analytically and the dome is the result of mostly competing order, similar to earlier work [41].

We can suggest a microscopic mechanism for the dome: as doping is lowered from the optimum $T_c$, the reduced number of localized bosons and free holes suppresses $T_c$. On the other hand, as doping is increased above optimum doping, the CDW energy which has allowed the formation of the stable, localized bosons drops precipitously. The virtual energy level which couples via Feshbach resonance also drops, thereby reducing $T_c$. It might even drop low enough that BCS-type superconductivity can complete, thereby missing the symmetries of the superconductivity and certainly changing certain properties. It is conceivable that all of the above mechanisms are utilized to some extent, but the one we have proposed seems to be the most fundamental.

**D. Pseudogap [43].**

The proposal that the pseudogap is "dimer" of CDW's/rotons is in a way a combination of two of the candidates proposed by others, namely a preformed pair and a vortex state. Certainly the description we have provided satisfies many of the experimental observation such as local diamagnetic supercurrents and especially the observations of Uemura and Kohsaka (from section 1). Through the simultaneous imaging of both r-space and k-space, the Bogoliubov quasiparticle region shrinks rapidly while transferring spectral weight to the higher energy r-space states which are not delocalized Cooper pairs, break translational and rotational symmetries at the atomic scale; these are the pseudogap states. Delocalized Cooper pairs vanish as the Mott insulator is approached by removing a few holes, leaving "localized" Cooper pairs.

**E. Nernst Effect.**

The observation of an extended Nernst signal above the critical temperature suggesting vortex excitations might be justified by an expected decomposition of a two-CDW/roton. The model presented here provides a natural explanation of the Nernst effect in a consistent theory for the entire superconducting dome.

**5. Summary, Conclusions and future work.**

We have proposed a normal state for the cuprates and fullerides which is both unusual and unique, namely paired fermions which are highly correlated and hence, act as a superconductor in many properties except off-diagonal long range order. The energy of binding on these pairs creates a virtual level which resonates with unbound holes (or electrons) to generate a delocalized Cooper pair and superconductivity. The stronger the energy of the virtual pair, the higher the transition temperature of the resulting superconductivity. We have also proposed a mechanism for the growth of the SC dome, the position of the peak, and the reduction of the dome.

We suspect that the model we have established might shed some light on other types of SC, such as the organic SC, the picntides, etc. based on their proximity in the Uemura diagrams. This is a topic we intend to explore as well as other suggested features of the model. In addition more study of the two-CDW/roton model seems appropriate.

**Appendix. Phase Rigidity (often called spin stiffness) in Superconductivity**

Superconductivity is a collective phenomenon, a so-called "emergent properties" where the whole is greater than the sum of parts. For example, a copper atom doesn't conduct electricity but a copper wire does as a result of collective effects. These effects are accomplished by a phase transition similar in certain respects to a liquid freezing to form a solid, where the solid phase is highly ordered, but the liquid is less so, if at all, and therefore has "broken translational symmetry". In addition the solid is rigid while the liquid is not. This notion of rigidity in a superconductor is more than the typical freezing of a liquid as can be illustrated by two additional properties: infinite conductivity (or persistent currents), and the expulsion of a magnetic field (Meissner-Ochsenfeld effect). Since the number of atoms, molecules, etc interacting is very large, the coherence which is responsible for the properties is macroscopic, not microscopic.

First, an experimental observation that establishes that a superconductor is a thermodynamic state: a superconducting material where the temperature is above $T_c$ can be taken by pathway 1) where the temperature is dropped below $T_c$, then a magnetic field applied, or 2) the opposite order which lead to the same state independent of path. The superconducting state expels the magnetic field from the sample, a property of thermal equilibrium called the Meissner-Ochsenfeld effect. The superconducting material is in a state which is a perfect diamagnet.

The collective mode called spin stiffness reflects a macroscopic order that produces the rigidity, analogous to a "universal XY model" which also describes a ferromagnet near the Curie point in the Stoner theory [44]. Both phenomena can be described by a complex order parameter having a form

$$\psi(\vec{r}) = |\psi| e^{i\theta(\vec{r})}$$

For $T > T_c$ the order parameter is zero, while $T < T_c$, then $\psi(T) \neq 0$. These features are part of the Ginzburg-Landau general model of bulk phase transitions, Figure x. So, when $T < T_c$ the direction of magnetization chooses one particular value of $\theta$. Heat the magnet above $T_c$ and then cool it again below $T_c$ will almost always produce a different angle $\theta$ for the direction of magnetization, but all magnets in the system will have this new angle $\theta$; this is spontaneous symmetry breaking. This same type of order will prevail over entropy at low enough temperature and produce phase rigidity.

The two key points that lead to the essential dynamical equations are 1) the phase and number operator are conjugate dynamical variables, i.e.

$$i\hbar \dot{N}_1 = \left[\hat{H}, N_1\right] = -i \partial \hat{H} / \partial \varphi_1$$

and the other Hamilton equation is

$$i\hbar \dot{\varphi}_1 = \left[\hat{H}, \varphi_1\right] = i \partial \hat{H} / \partial N_1$$

These two equations, applicable to superconductivity and superfluidity, originated with London [39], the second one illustrating the response of the system to forces. They apply both to operators and mean values of the wave packet phase. For an isolated system the Hamiltonian commutes with the number operator and is independent of total phase and obeys gauge invariance as usual. The number of particles entering or leaving an isolated system is zero.

Consider two separate superconducting (or superfluid) systems connected by a "bridge" that has the same superconducting properties. Then, there is a dependency of the energy on the phase difference of the two sides,

$$U = U(\varphi_1 - \varphi_2)$$

Supercurrents are then possible between the two sides

$$\frac{dN_1}{dt} = -\frac{dN_2}{dt} = \frac{1}{\hbar}\frac{\partial U(\varphi_1 - \varphi_2)}{\partial(\varphi_1 - \varphi_2)}$$

The energy as a function of the gradient of the phase is a minimum when the gradient is zero and assumes a quadratic form for small gradient values, so the general form is

$$U = \int_V d^3\vec{r}\, U(\vec{r}) = \frac{\hbar^2}{2m} \int_V d^3\vec{r}\, n_s (\nabla\varphi)^2$$

where $n_s$ is equal to the electron pairs or total number of particles in the system, i.e. the superfluid density. Then, the supercurrent density is the derivative of U with respect to the gradient of the phase

$$\vec{j}_s = \frac{1}{\hbar}\frac{dU}{d(\vec{\nabla}\varphi)} = \frac{\hbar}{2m} n_s \vec{\nabla}\varphi$$

This is completely analogous to the XY ferromagnet when this alignment can be described as

$$+J\left|\vec{\nabla}\cdot\vec{S}\right|^2 \quad (\text{or } \sum_{\langle i,j \rangle} -\vec{s}_i \cdot \vec{s}_j\,)$$

so we can write a similar relationship for our phase rigidity as

$$H = \frac{1}{2} n_s (\vec{\nabla}\varphi)^2$$

where $n_s$ is the "stiffness" and $\vec{j} = n_s \vec{\nabla}\theta$ is the current density (note $\rho_s$ is also used in the literature). These last two equations lend some credibility to the general integrated form,

$$H = \frac{1}{2} n_s (\nabla\varphi + q\vec{A})^2$$

which explicitly shows the U(1) phase of the order parameter of the charged condensate, q, in the presence of an external field with vector potential $\vec{A}$.

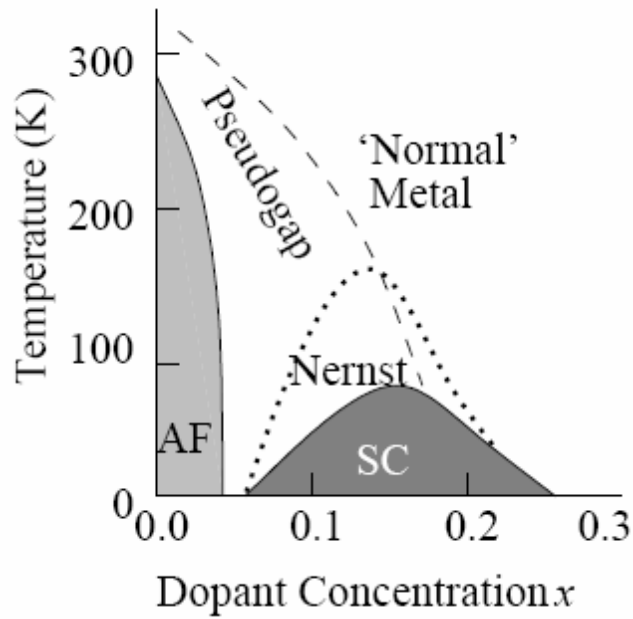

Figure 1 Generic phase diagram of cuprate superconductors. The SC dome has three components: an initial rise generated from antiferromagnetism by doping and controlled by phase coherence; the second "peak" controlled by the Feshbach resonance of the mobile spins with the virtual bound level of the boson-like "local" SC, and lastly, the diminishing segment where various order parameters interact. These include the disappearing "local: SC responsible for the pseudogap and the Nernst effect, uneven charge build-up on the various cuprate layers and ultimately a regeneration of BCS SC. The rich phase diagram may also include two quantum critical points at about 0.05 and 0.23 doping levels.

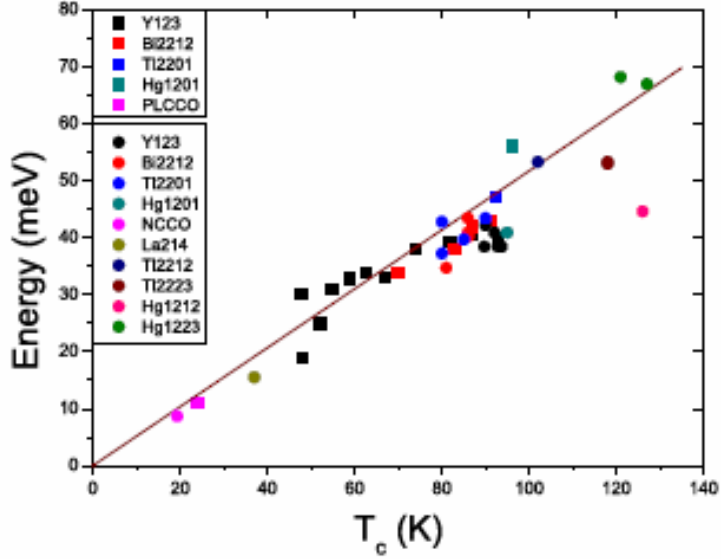

Figure 2. Plot of inelastic neutron scattering (INS - solid squares) and Raman electronic scattering (ERS - solid circles) of the $A_{1g}$ mode. The INS data corresponds to the $\pi - \pi$ resonance energy, a singlet-triplet excitations while the ERS represents a singlet mode. We will argue that these represent the correlated charge attribute of the charged roton-like structure in resonance with two spin density waves and suggest that the formula $k_B T_c = \dfrac{E_g}{\kappa}$ relates the energy $E_g$ to the depth of the bound state in the Feshbach resonance (next figure) which is the stability of the superconducting Cooper pair.

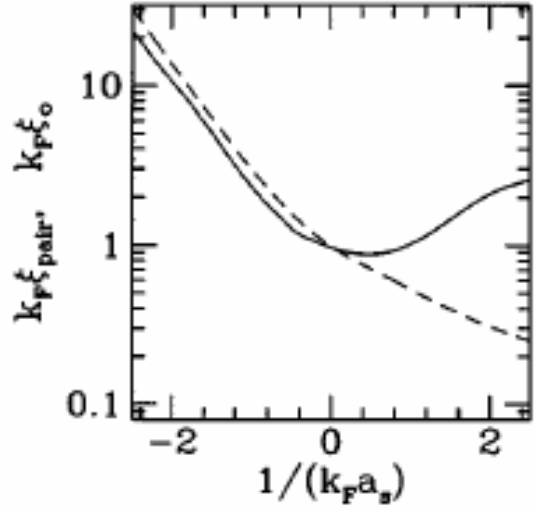

Fig 3 The Ginsburg-Landau (GL) coherence length $\xi_0$ is the solid line (units of $k_F^{-1}$) and the pair size $\xi_{pair}$ is the dashed line plotted as a function of the coupling $1/k_F a_s$. Here the working definition of pair size is $\xi_{pair}^2 = -\langle \psi_k | \nabla_k^2 | \psi_k \rangle / \langle \psi_k | \psi_k \rangle$ and $\psi_k = \Delta_0 / 2E_k$ is the T = 0 pair wave function (after Engelbrecht et al [45]).

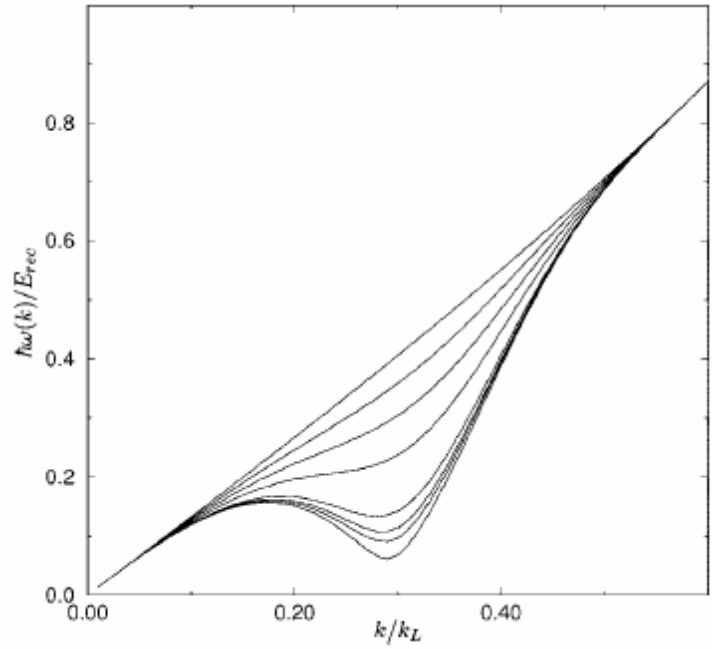

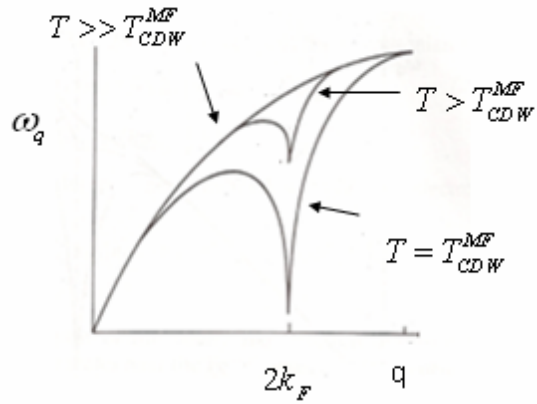

Figure 4. Comparison of roton formation [19] with the formation of the charge density wave as the temperature is lowered (Kohn anomaly).

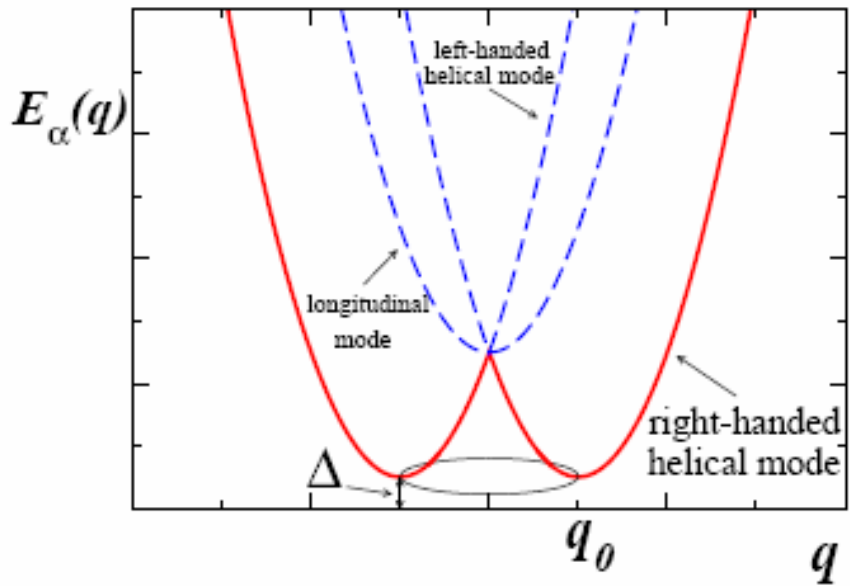

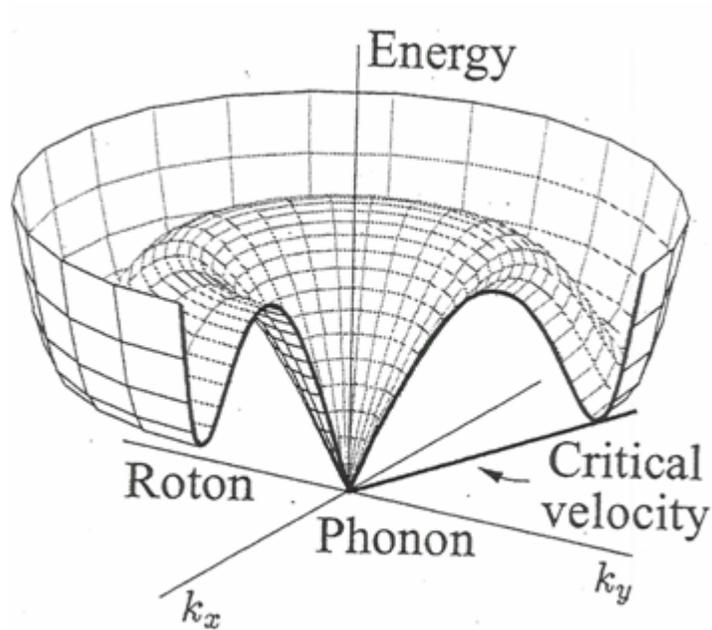

Figure 5 Comparison of the potential creating a circular CDW and a roton. Both are symmetry breaking and seem to have separate origins, but despite this they can stabilize highly correlated two bound-rotons and two bound-CDW's resulting in "local" superconductivity.

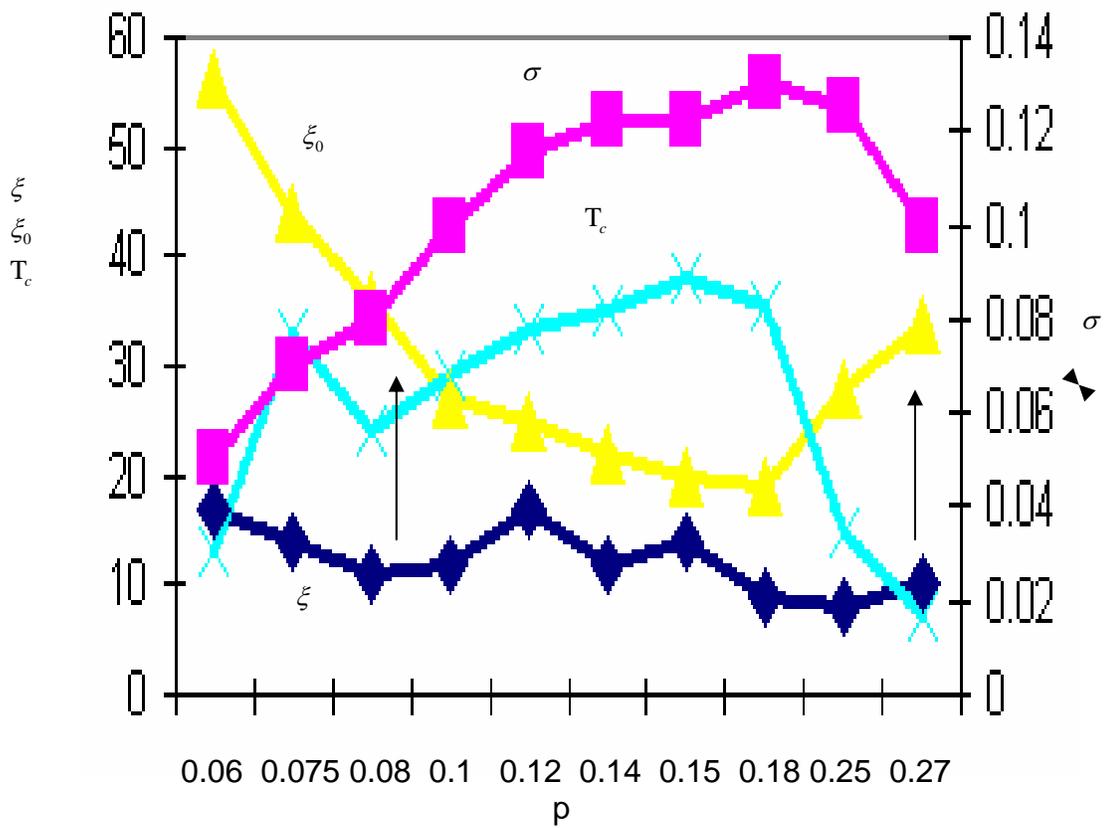

Figure 6. Concurrent graphs of $\xi$, spin correlation length, $\xi_0$, Cooper pair correlation length, $T_c$, and $\delta$, extrapolated values of the incommensurability. As we suggest in the manuscript, resonance is lost at low and high doping (arrows), through for different reasons.

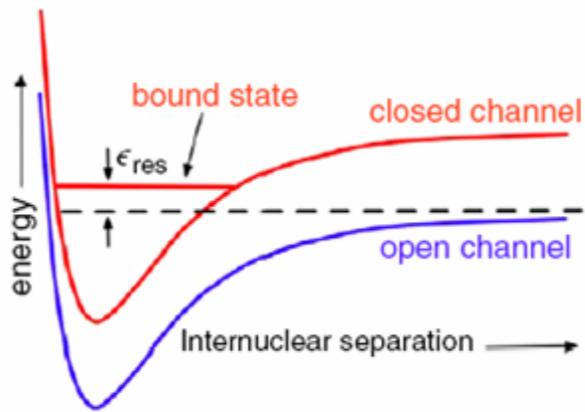

Fig 7 Illustration of a Feshbach resonance where a bound state energy $\left(\varepsilon_{res}=2\nu\right)$ is in resonance with the scattering threshold of two Fermions in an open channel. The energy $\varepsilon_{res}$ is the energy axis in Fig 2 and is the main component of $T_c$. The bound state can be influenced by specific nature of the material, such as Josephson energies and charge layers modifying the binding energy of the Cooper pair and hence, $T_c$.